\documentclass[lettersize,journal]{IEEEtran}
\usepackage{amsmath,amsfonts}
\usepackage{algorithmic}
\usepackage{algorithm}
\usepackage{array}
\usepackage[caption=false,font=normalsize,labelfont=sf,textfont=sf]{subfig}
\usepackage{textcomp}
\usepackage{stfloats}
\usepackage{url}
\usepackage{verbatim}
\usepackage{graphicx}
\usepackage{cite}

\begin{document}
\title{Theoretical Analysis of Heterodyne Rydberg Atomic Receiver Sensitivity Based on Transit Relaxation Effect and Frequency Detuning}

\author{Shanchi Wu, Chen Gong, Shangbin Li, Rui Ni, Jinkang Zhu
	\thanks{This work was supported in part by National Natural Science Foundation of China under Grant 62171428, in part by Key Research Program of Frontier Sciences of CAS under Grant QYZDY-SSW-JSC003, and Huawei Innovation Project.}	
	\thanks{Shanchi Wu, Chen Gong, Shangbin Li, and Jinkang Zhu are with University of Science and Technology of China, Email address: wsc0807@mail.ustc.edu.cn,\{cgong821, shbli, jkzhu\}@ustc.edu.cn.}
	\thanks{Rui Ni is with Huawei Technology, Email address: raney.nirui@huawei.com.}}

\maketitle

\begin{abstract} 
We conduct a theoretical investigation into the impacts of local microwave electric field frequency detuning, laser frequency detuning, and transit relaxation rate on enhancing heterodyne Rydberg atomic receiver sensitivity. To optimize the output signal's amplitude given the input microwave signal, we derive the steady-state solutions of the atomic density matrix. Numerical results show that laser frequency detuning and local microwave electric field frequency detuning can improve the system detection sensitivity, which can help the system achieve extra sensitivity gain. It also shows that the heterodyne Rydberg atomic receiver can detect weak microwave signals continuously over a wide frequency range with the same sensitivity or even more sensitivity than the resonance case. To evaluate the transit relaxation effect, a modified Liouville equation is used. We find that the transition relaxation rate increases the time it takes to reach steady state and decreases the sensitivity of the system detection.
\end{abstract}


\section{Introduction}
\IEEEPARstart{R}{ydberg}  atoms show extremely strong microwave transition electric dipole moments, which are sensitive to external electromagnetic fields. At room temperature, electromagnetic fields can be measured with high sensitivity and precision using atomic quantum coherence effects.

Utilizing electromagnetically induced transparency (EIT) spectroscopy, a Rydberg atomic sensor with sensitivity 30 uV$\cdot$cm$^{-1}$Hz$^{-1/2}$ and minimum detectable electric field intensity of $8$ uV/cm has been demonstrated \cite{sedlacek2012microwave}. Rydberg atomic sensors' sensitivity can be raised to $2$ uV$\cdot$cm$^{-1}$Hz$^{-1/2}$ using balanced homodyne detection and frequency modulation methods based on optical interferometers \cite{kumar2017atom, kumar2017rydberg}. Combining the Rydberg atomic sensor with the traditional superheterodyne approach results in a significant increase in sensitivity. Such method introduces a local microwave signal and the phase and frequency measurement of microwaves can be realized by the Rydberg atomic sensor, improving the sensitivity of microwave electric field detection to $55$ nV$\cdot$cm$^{-1}$Hz$^{-1/2}$ \cite{jing2020atomic}. The setting of laser parameters also affects the electric field detection sensitivity of the Rydberg atomic sensor. Numerical simulations and experimental results show that the sensitivity of the Rydberg atomic sensor is related to the amplitude intensity of the two lasers and microwave electric. The microwave electric detection sensitivity can be improved to $12.5$ nV$\cdot$cm$^{-1}$Hz$^{-1/2}$ by detuning the coupling laser frequency, which is the highest sensitivity achieved in experiments so far \cite{cai2022sensitivity}. For applications in communication systems, predicting the Rydberg atomic sensor performance under different parameter settings through theoretical analysis and numerical simulation is significant.

Rydberg atomic sensors' detection sensitivity is fundamentally constrained by quantum noise, which is orders of magnitude less prevalent than thermal noise. The coherence time of the atoms can be decreased in real systems by a variety of variables, which lowers the system's sensitivity to detection. The main relaxation mechanisms that affect Rydberg atomic sensor's sensitivity include collision broadening, transition time broadening \cite{sagle1996measurement}, power broadening \cite{sautenkov2017power} and laser linewidth \cite{lu1997electromagnetically, tanasittikosol2011microwave}. Atomic collisions with cavity walls as well as atomic collisions with one another cause the collision broadening effect. The collision broadening effect can be efficiently decreased by low atomic density. The atoms' mobility causes the transition time broadening effect, which depends on the temperature, atomic mass, and laser beam size. The power broadening originates from the instability of laser power, which can be squeezed by the power stabilization module in experiment. The laser linewidth is an inherent property of lasers, and narrow linewidth laser can improve the performance of Rydberg atomic sensors. Finding technical methods to eliminate or reduce the influence of these factors is crucial for future applications in communication system.

Recent progress in experiments show that reducing the laser linewidth and power fluctuation are feasible improvement methods. Experimental techniques with ultrastable cavities can achieve extremely narrow linewidth laser beam and high power stabilization, which can increase the system sensitivity to the level higher than $100$ nV$\cdot$cm$^{-1}$Hz$^{-1/2}$. The optical readout noise is the dominant factor limiting the detection sensitivity \cite{jing2020atomic, cai2022sensitivity}. Except for apparatus performance improvement, new detection and readout techniques are the options left. In the perspective of readout techniques, the experiment based on optical interferometer provides ideas for reducing readout noise. Methods that combine circulating cavity with compressed state light have the potential to further reduce the readout noise limit of the system. Existing experimental schemes adopt different detection techniques depending on the atomic response regime, such as AT splitting effect of atomic resonance \cite{sedlacek2012microwave}, AC Stark effect under off-resonant \cite{anderson2021self}, and the heterodyne readout technique with tuning capability \cite{gordon2019weak}. A continuously tunable electric field measurement based on the far off-resonant AC stark effect in a Rydberg atomic vapor cell shows comparable detection sensitivity with a resonant microwave-dressed Rydberg heterodyne receiver using the same system \cite{simons2021continuous}. Additional research about electric field detuning can make smooth transitions among these operating regimes, especially for modulated signals. The frequency modulation spectroscopy \cite{kumar2017rydberg}, lower noise photodetector, and the setting of operating state in the experiment are all sensitivity optimization methods worth trying.

In our preliminary work, we have evaluated the optimal value of local microwave frequency detuning under high transmittance case \cite{wu2022rydberg}. In this work, we further characterize the effects of local microwave, probing laser and coupling laser frequency detuning on the sensitivity of Rydberg atomic sensors. In addition, we investigate the impact of transit relaxation on the sensitivity of the system. Assume that the heterodyne scheme can detect the microwave signal amplitude, phase, and frequency, and the system can be tuned by changing the local microwave signal frequency. We aim to build on this heterodyne reception framework. 

The reminder of this paper is organized as follows. In Section II, we outline the basic detection principles of heterodyne schemes. In Section III, we optimize the frequency parameters based on atomic four-level model. We theoretically investigate the detection response to the local microwave, coupling laser and probing laser frequency detuning. In Section IV, we consider the transit relaxation effect caused by the atomic motion, and numerically characterize the system sensitivity variation. Finally in Section V, we conclude this work.

\section{Theoretical Model}

\begin{figure*}[htbp]
	\centering
	\includegraphics[width = 5in]{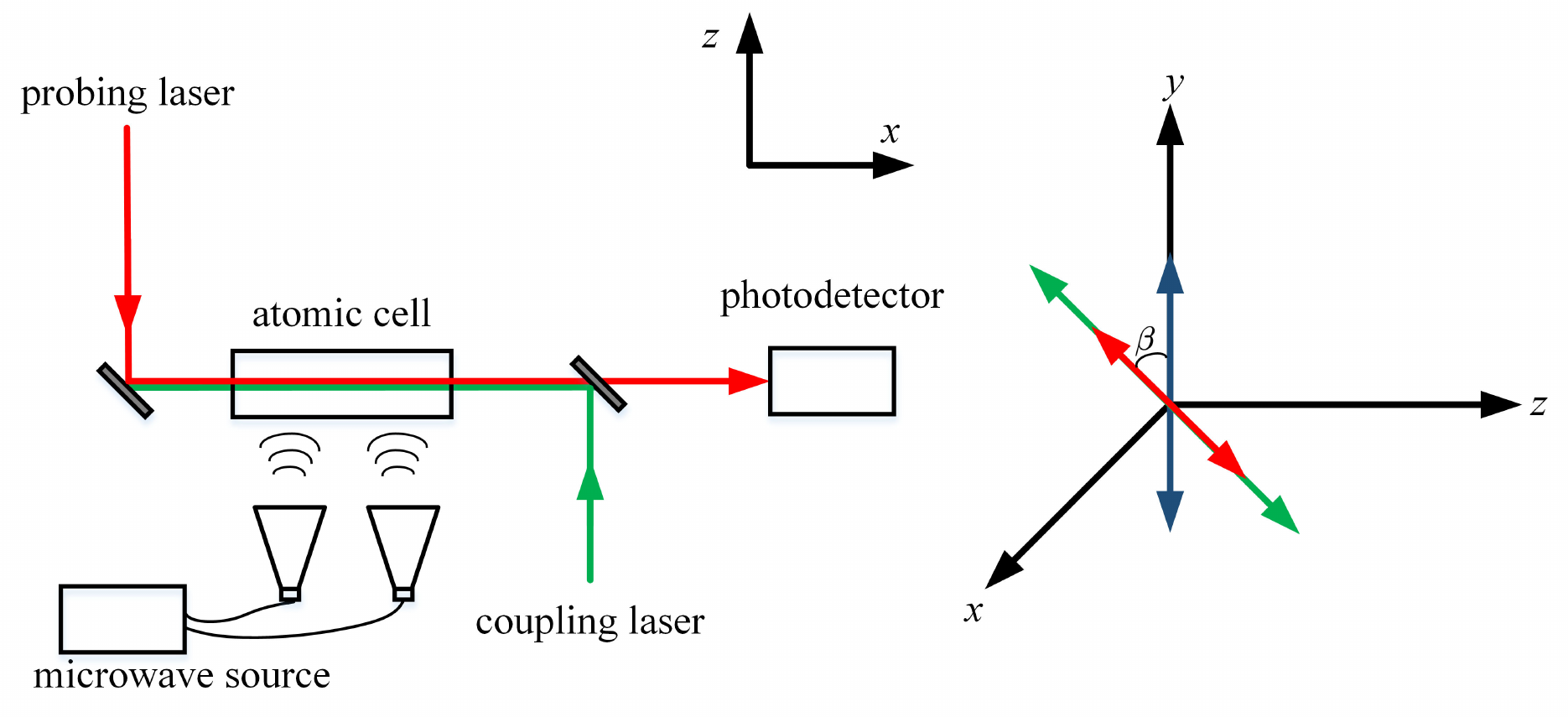}
	\caption{The diagram of the atomic heterodyne detection system.}
	\label{fig.system}
\end{figure*}

A schematic diagram of the detection system based on Rydberg atomic sensor is shown in Fig. \ref{fig.system}. The probing laser and coupling laser simultaneously excite the atoms in the atomic vapor cell, and the photodetector receives the transmission probing laser. The probing laser and coupling laser are locked on the EIT transmission resonance. When the microwave signal radiates cross the atomic vapor cell, the resonant microwave electric field can help improve the transmission performance and the AC Stark shift caused by off-resonant microwave electric field also changes the probing laser transmission, thereby changing the signal intensity received by the photodetector. Because the system response is sensitive to the direction of polarization, we set that the system lies in the $xoz$ plane, and the probing laser and coupling laser have the same polarization in the $yoz$ plane. The incident microwave electric filed has the $y$-axis polarization and $\beta$ is the angle difference. It has been demonstrated that the microwave electric field polarization can also be obtained by atomic system \cite{sedlacek2013atom}. In this work, we assume that the lasers and microwave signal have the same polarization, i.e., the polarization angle $\beta$ between lasers and microwave is zero.

\begin{figure}[htbp]
	\centering
	\includegraphics[width = 2.5in]{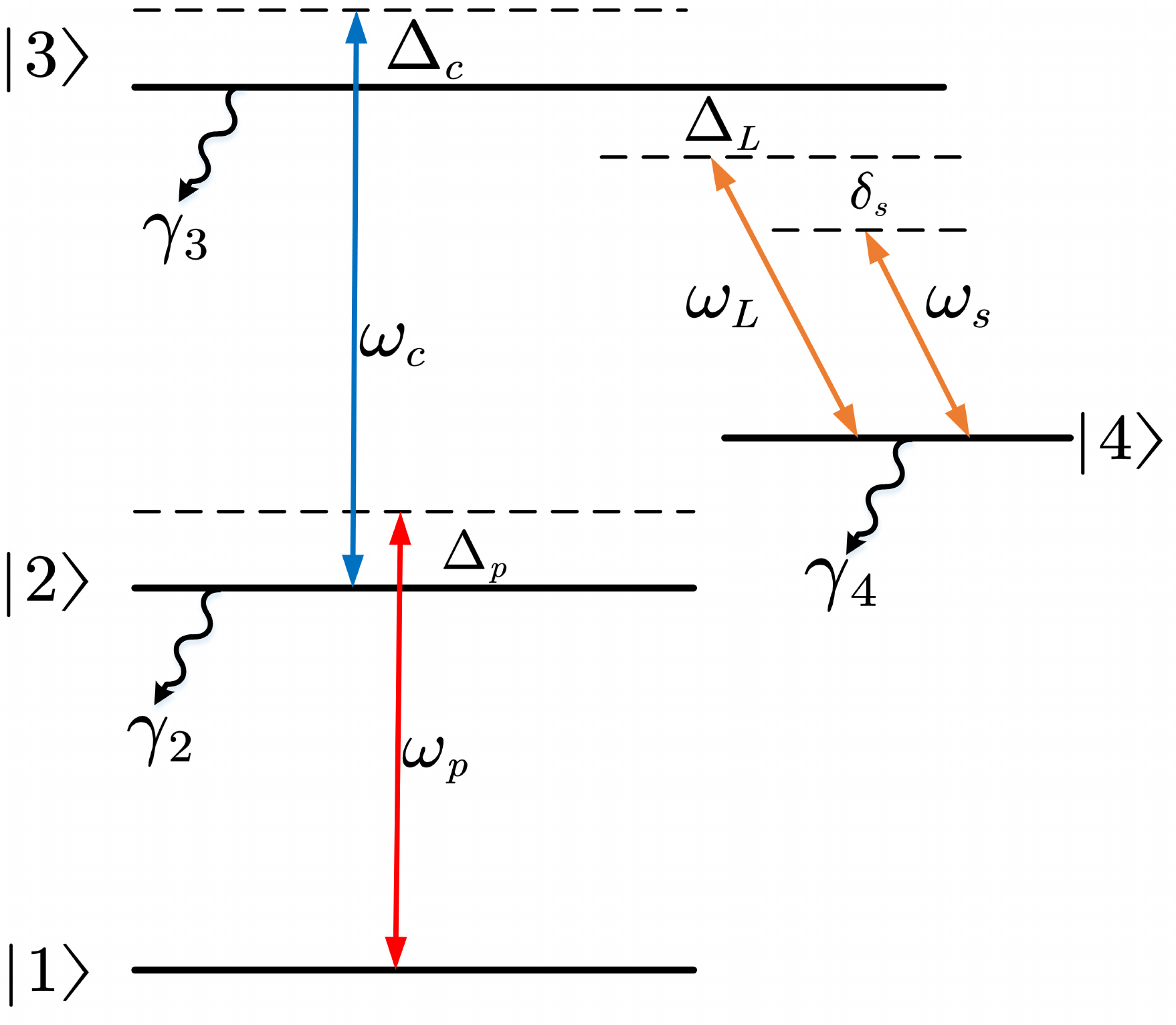}
	\caption{The diagram of four-level configuration.}
	\label{fig.level}
\end{figure}

The diagram for a typical atomic four-level structure is shown in Fig. \ref{fig.level}. The probing laser and coupling laser are tuned with detuning $\Delta _p$ and $\Delta _c$, respectively; and the local microwave electric field is tuned with detuning $\Delta _L$. Assume that the frequency and phase difference between local microwave electric field and signal microwave electric field is $\delta _s$ and $\phi_s$, respectively. The corresponding Hamiltonian in the rotating frame is given by
\begin{equation}
H=\frac{{\hbar }}{2}\left( \begin{matrix}
	0&		\Omega _p&		0&		0\\
	\Omega _p&		2\Delta _p&		\Omega _c&		0\\
	0&		\Omega _c&		2\left( \Delta _p+\Delta _c \right)&		\Omega _L+\Omega _se^{-iS\left( t \right)}\\
	0&		0&		\Omega _L+\Omega _se^{iS\left( t \right)}&		2\left( \Delta _p+\Delta _c+\Delta _L \right)\\
\end{matrix} \right) ,
\end{equation} 
where $\Omega _{p}$ and $\Omega _{c}$ are the corresponding Rabi frequencies for transitions $\left. |1 \right> \rightarrow \left. |2 \right>$ and $ \left. |2 \right> \rightarrow \left. |3 \right>$, respectively. The local microwave electric field and signal microwave electric field couple with Rydberg transition between state $ \left. |3 \right> $ and state $ \left. |4 \right>$, with Rabi frequencies $\Omega_L$ and $\Omega_s$, respectively. Let $\Omega =|\Omega _L+\Omega_se^{-iS\left( t \right)}|$, where $S\left( t \right) =2\pi \delta _st+\phi _s$ is the cumulative phase difference of the signal microwave relative to the local microwave.

Due to the relaxation effect caused by the spontaneous radiation, collision of atoms and transit relaxation effect, as well as the atoms repopulation, the complete Liouville equation for the rotating-frame density matrix of the system can be written as \cite{auzinsh2010optically}
\begin{equation}
	i{\hbar }\frac{d}{dt}\rho =\left[ H,\ \rho \right] -i{\hbar }\frac{1}{2}\left( \Gamma \rho +\rho \Gamma \right) +i{\hbar }\Lambda,
\end{equation}
where $\rho$ is the density matrix. Relaxation matrix $\Gamma$, which represents the relaxation rate of each state, is given by
\begin{equation}
	\Gamma =\left( \begin{matrix}
		\gamma&		0&		0&		0\\
		0&		\gamma +\gamma _2&		0&		0\\
		0&		0&		\gamma +\gamma _3+\gamma _c&		0\\
		0&		0&		0&		\gamma +\gamma _4\\
	\end{matrix} \right),
\end{equation}
where $\gamma_{2}$, $\gamma_{3}$ and $\gamma_{4}$ represent the spontaneous decay rates of atoms on the three high levels; $\gamma_c$ and $\gamma$ are the relaxation rates of the atoms collision and transit effect, respectively. We assume that each level undergoes the same relaxation rate $\gamma$ due to the exit of atoms from the laser beam.

In addition, the atoms that spontaneously decay from the upper states also repopulate the lower sates. In this work, we only consider the decay paths shown in Fig. \ref{fig.level}. Repopulation matrix $\Lambda$ is given by
\begin{equation}
	\Lambda =\left( \begin{matrix}
		\gamma +\gamma _2\rho _{22}+\gamma _4\rho _{44}&		0&		0&		0\\
		0&		\gamma _3\rho _{33}&		0&		0\\
		0&		0&		0&		0\\
		0&		0&		0&		0\\
	\end{matrix} \right).
\end{equation}
The dynamics equation of the atomic system can be expressed as
\begin{equation}\label{eq.basic}
	\frac{d}{dt}\rho =-\frac{i}{{\hbar }}\left[ H,\rho \right] +L,
\end{equation}
where $L=-\frac{1}{2}\left( \Gamma \rho +\rho \Gamma \right) + \Lambda$ represents the Lindbladian operator that defines the relaxation progress in the system.

Under resonant conditions, i.e., $\Delta _c=\Delta _p=\Delta _L=0$, ignoring the atom collision and transit time broadening effect, we have $\gamma=\gamma_c=0$. The steady-state solution $\rho_{21}(t)$ is given by
\begin{equation}\label{eq.steady}
	\rho _{21}(t)=-\frac{i\gamma _2\Omega _p\Omega ^2}{\gamma _{2}^{2}\Omega ^2+2\Omega _{c}^{2}\Omega _{p}^{2}+2\Omega _{p}^{4}+2\Omega _{p}^{2}\Omega ^2}.
\end{equation}
The linear susceptibility can be written as
\begin{equation}\label{eq.chi}
	\chi \left( t \right) =-\frac{2N_0\mu _{12}^{2}}{{\hbar }\epsilon _0\Omega _p}\rho _{21}\left( t \right),
\end{equation}
where $N_0$ is the total density of atoms, $\mu_{12}$ is the dipole moment of transition $|1\rangle \rightarrow |2\rangle$ and $\epsilon_0$ is the permittivity in vacuum. Assuming $\Omega_s \ll \Omega_L$, it has been proved that the optimal local microwave filed amplitude can improve the system sensitivity. The optimal Rabi frequency of $\Omega_L$ is given by \cite{jing2020atomic}
\begin{equation}\label{eq.omegaL}
	\Omega _L=\Omega _p\sqrt{\frac{2\left( \Omega _{c}^{2}+\Omega _{p}^{2} \right)}{3\left( 2\Omega _{p}^{2}+\gamma _{2}^{2} \right)}}.
\end{equation}
 
In this work, based on the steady-state solution of the system Liouville equation, we derive the expression of the linear susceptibility and output optical power, and then find the optimal values. In general, we expect that the output optical power has the formalism as follows,
\begin{equation}\label{eq.output}
	P\left( t \right) =\bar{P}_0+\kappa \Omega _s\cos \left( 2\pi \delta _st+\phi _s \right).
\end{equation}
We aim to maximize the absolute value of conversion coefficient $\kappa$, which represents the detection sensitivity.

\section{Parameters Optimization}\label{sec.frequency}
\subsection{Methodology}
In experimental systems, in order to reduce the interaction and collision rate between atoms, the atomic density and laser power are usually limited, although increasing the laser power can theoretically increase the signal-to-noise ratio of the output signal and improve the sensitivity. Therefore, in the current experimental setups, the detection laser power is usually at a weak level, and the coupled laser power is also limited. For Rydberg atom sensors using EIT readout, the optimal Rabi frequencies for transitions $|1\rangle \rightarrow |2\rangle$ and $|2\rangle \rightarrow |3\rangle$ depend on the decay rate of each state and relaxation rate caused by atom collision \cite{meyer2021optimal}. We analyze the influence of laser frequencies and local microwave electric field frequency detuning on the system sensitivity.

To gain an intuitive understanding, we begin with analytical derivations. We set a detuning parameter $ \Delta$ (which can be $\Delta_p$, $\Delta _c$, or $\Delta _L$). Assuming $\gamma _3=\gamma _4=\gamma_c=\gamma=0$, the steady-state solution $\rho _{21}\left( \Delta \right)$ can be obtained based on Eq. (\ref{eq.basic}). The corresponding linear susceptibility of the atomic system can be written as
\begin{equation}
\mathfrak{J}\left( \chi \left( t,\ \Delta \right) \right) =\chi _0\left( \Delta \right) +\chi _1\left( \Delta \right) \Omega _s\cos \left( 2\pi \delta _st+\phi _s \right),
\end{equation}
which has a time-invariant part $\chi _0\left( \Delta \right)$ and the coefficient of time variant part $\chi _1\left( \Delta \right)$. The first part contributes a constant signal component at the photodetector, and the second part reflects the information transmission of signal microwave field. 

The probing laser transmission is associated with the imaginary component of the susceptibility as
\begin{equation}
	\begin{aligned}
		P\left( t \right) &=P_ie^{-kL\Im \left( \chi \left( t,\ \Delta \right) \right)}\\
		&=e^{-kL\chi _0\left( \Delta \right)}P_ie^{-kL\chi _1\left( \Delta \right) \Omega _s\cos \left( 2\pi \delta _st+\phi _s \right)},
	\end{aligned}
\end{equation}
where $k=2\pi /\lambda _p$ is the wavevector of probing laser, and $L$ is the length of atomic vapor cell along the laser beam propagation direction. The peak-to-peak value of the output signal is
\begin{equation}\label{eq.12}
P_{pp}\left( \Delta \right) =e^{-kL\chi _0\left( \Delta \right)}P_i\left( e^{kL\chi _1\left( \Delta \right) \Omega _s}-e^{-kL\chi _1\left( \Delta \right) \Omega _s} \right).
\end{equation}
Since the amplitude of microwave signal is linearly dependent on its corresponding Rabi frequency $ \Omega_s$, we can optimize coefficients $\chi _0\left( \Delta \right)$ and $ \chi _1\left( \Delta \right) $ in Eq. (\ref{eq.12}) to maximize $P_{pp}\left( \Delta \right)$. Two cases are considered in the following.

\textbf{General case:} In weak signal regime, $|kL\chi _1\left( \Delta \right) \Omega _s|$ is much lower than one. The output laser power can be approximated as
\begin{equation}
	\begin{aligned} 	
		P\left( t \right) &\approx e^{-kL\chi _0\left( \Delta \right)}P_i\left( 1-kL\chi _1\left( \Delta \right) \Omega _s\cos \left( 2\pi \delta _st+\phi _s \right) \right)\\ 	
		&= \bar{P}_0\left( \Delta \right) +\kappa \left( \Delta \right) P_i\Omega _s\cos \left( 2\pi \delta _st+\phi _s \right)\\ 
	\end{aligned},
\end{equation}
where $\bar{P}_0\left( \Delta \right) =P_ie^{-kL\chi _0\left( \Delta \right)}$ is the DC component of the output optical power and $\kappa \left( \Delta \right) =-e^{-kL\chi _0\left( \Delta \right)}kL\chi _1\left( \Delta \right)$ represents the conversion coefficient to signal. The DC component is associated with the photodetector operation point. Our goal is to maximize the conversion coefficient $|\kappa(\Delta)| $ through parameter optimization.

\textbf{High transmittance case:} If $|kL\chi _0\left( \Delta \right)|$ is much lower than one, $e^{-kL\chi _0\left( \Delta \right)} $ can be approximated as a constant. The output laser power can be approximated as
\begin{equation}
	\begin{aligned} 	
		P\left( t \right) 
		&\approx P_i\left( 1-kL\chi _1\left( \Delta \right) \Omega _s\cos \left( 2\pi \delta _st+\phi _s \right) \right)\\ 	
		&= \bar{P}_{0}^{'}\left( \Delta \right) +\kappa '\left( \Delta \right) P_i\Omega _s\cos \left( 2\pi \delta _st+\phi _s \right),\\ 
	\end{aligned}
\end{equation}
where $\bar{P}_{0}^{'}\left( \Delta \right) \approx P_i$ is the DC component of the output optical power and $\kappa '\left( \Delta \right) =-kL\chi _1\left( \Delta \right)$ represents the conversion coefficient. In high transmittance case, the optimal problem of maximizing conversion coefficient $|\kappa '(\Delta)|$ is equivalent to maximizing the coefficient $|\chi _1|$.

\subsection{Local Microwave Detuning}\label{sec.l_detuning}

Assuming that $\Delta _c=\Delta _p=\gamma _3=\gamma _4=0$, the steady-state solution $\rho_{21}(\Delta_{L})$ can be obtained according to Eq. (\ref{eq.basic}) as Eq. (\ref{eq.rho21_L}).
\begin{figure*}
\begin{equation}\label{eq.rho21_L}
	\begin{aligned}
		\rho _{21}( \Delta _L )&=-\frac{i\Omega _p\Omega ^2\left( \gamma _2\Omega ^2+2i\Delta _L\Omega _{c}^{2} \right)}{\gamma _{2}^{2}\Omega ^4+4\Delta _{L}^{2}\Omega _{c}^{4}+8\Delta _{L}^{2}\Omega _{c}^{2}\Omega _{p}^{2}+4\Delta _{L}^{2}\Omega _{p}^{4}+2\Omega _{c}^{2}\Omega _{p}^{2}\Omega ^2+2\Omega _{p}^{4}\Omega ^2+2\Omega _{p}^{2}\Omega ^4}\\
		&=-\frac{i\gamma _2\Omega _p\Omega ^4-2\Delta _L\Omega _p\Omega ^2\Omega _{c}^{2}}{\gamma _{2}^{2}\Omega ^4+4\Delta _{L}^{2}\left( \Omega _{c}^{2}+\Omega _{p}^{2} \right) ^2+2\Omega ^2\Omega _{p}^{2}\left( \Omega _{c}^{2}+\Omega _{p}^{2}+\Omega ^2 \right)}.\\
	\end{aligned}
\end{equation}
\end{figure*}
In the weak signal regime, $\Omega_s \ll \Omega_L$, the first order approximation of $\Omega_s / \Omega_L$ is given by Eq. (\ref{eq.rho21_L_Im}).
\begin{figure*}
\begin{equation}\label{eq.rho21_L_Im}
	\begin{aligned}
		\Im \left( \rho _{21}( \Delta _L ) \right) \approx -\frac{\gamma _2\Omega _p\Omega _{L}^{4}}{\gamma _{2}^{2}\Omega _{L}^{4}+4\Delta _{L}^{2}\left( \Omega _{c}^{2}+\Omega _{p}^{2} \right) ^2+2\Omega _{L}^{2}\Omega _{p}^{2}\left( \Omega _{p}^{2}+\Omega _{c}^{2}+\Omega _{L}^{2} \right)}\\
		-\frac{4\gamma _2\Omega _{L}^{3}\Omega _p\left( \Omega _{p}^{2}+\Omega _{c}^{2} \right) \left( \Omega _{L}^{2}\Omega _{p}^{2}+4\Delta _{L}^{2}\left( \Omega _{p}^{2}+\Omega _{c}^{2} \right) \right)}{\left[ \gamma _{2}^{2}\Omega _{L}^{4}+4\Delta _{L}^{2}\left( \Omega _{c}^{2}+\Omega _{p}^{2} \right) ^2+2\Omega _{L}^{2}\Omega _{p}^{2}\left( \Omega _{c}^{2}+\Omega _{p}^{2}+\Omega _{s}^{2} \right) \right] ^2}\Omega _s \cos S\left( t \right).\\
	\end{aligned}
\end{equation}
\end{figure*}
The corresponding linear susceptibility of the atomic system can be written as
\begin{equation}
	\Im \left( \chi \left( t,\ \Delta _L \right) \right) =\chi _0\left( \Delta _L \right) +\chi _1\left( \Delta _L \right) \Omega _s\cos S\left( t \right),
\end{equation}
which has a time-invariant part as Eq. (\ref{eq.chi0}),
\begin{figure*}
\begin{equation}\label{eq.chi0}
	\chi _0\left( \Delta _L \right) =\frac{2N_0\mu _{12}^{2}}{{\hbar }\epsilon _0}\frac{\gamma _2\Omega _{L}^{4}}{\gamma _{2}^{2}\Omega _{L}^{4}+4\Delta _{L}^{2}\left( \Omega _{c}^{2}+\Omega _{p}^{2} \right) ^2+2\Omega _{L}^{2}\Omega _{p}^{2}\left( \Omega _{p}^{2}+\Omega _{c}^{2}+\Omega _{L}^{2} \right)},
\end{equation}
\end{figure*}
and a time variant part coefficient as Eq. (\ref{eq.chi1}).
\begin{figure*}
\begin{equation}\label{eq.chi1}
	\chi _1\left( \Delta _L \right) =\frac{2N_0\mu _{12}^{2}}{{\hbar }\epsilon _0}\frac{4\gamma _2\Omega _{L}^{3}\left( \Omega _{p}^{2}+\Omega _{c}^{2} \right) \left( \Omega _{L}^{2}\Omega _{p}^{2}+4\Delta _{L}^{2}\left( \Omega _{p}^{2}+\Omega _{c}^{2} \right) \right)}{\left[ \gamma _{2}^{2}\Omega _{L}^{4}+4\Delta _{L}^{2}\left( \Omega _{c}^{2}+\Omega _{p}^{2} \right) ^2+2\Omega _{L}^{2}\Omega _{p}^{2}\left( \Omega _{c}^{2}+\Omega _{p}^{2}+\Omega _{L}^{2} \right) \right] ^2}.
\end{equation}
\end{figure*}

In the general case, the sensitivity optimization problem is formulated as follows.

\textbf{(P1): Sensitivity Maximization in General Case via Local Microwave Detuning}
\begin{equation}\label{eq.opti1}
	\Delta _{L}^{*}=\underset{\Delta _L}{\text{argmax }}\kappa(\Delta_L)=\underset{\Delta _L}{\text{argmax }} e^{-kL\chi _0\left( \Delta _L \right)} \chi _1\left( \Delta _L \right).
\end{equation}

Adopting the following variable substitution,
\begin{equation}\label{eq.sub}
	\frac{\Omega _{p}^{2}}{\gamma _{2}^{2}}\rightarrow x,\ \frac{\Omega _{c}^{2}}{\gamma _{2}^{2}}\rightarrow y,\,\,\frac{\Omega _{L}^{2}}{\gamma _{2}^{2}}\rightarrow z,\,\,\frac{\Delta _{L}^{2}}{\gamma _{2}^{2}}\rightarrow w,
\end{equation}
and we can get an equivalent optimization problem as Eq. (\ref{eq.opti_l1}),
\begin{figure*}
\begin{equation}\label{eq.opti_l1}
w^*=\underset{w}{\text{arg}\max}\left\{ \begin{array}{c}
	\exp[-\frac{C z^2}{z^2+4w\left( x+y \right) ^2+2xz\left( x+y+z \right)}]\cdot \frac{z^{3/2}\left( x+y \right) \left( xz+4w\left( x+y \right) \right)}{\left[ z^2+4w\left( x+y \right) ^2+2xz\left( x+y+z \right) \right] ^2}\\
\end{array} \right\},
\end{equation}
\end{figure*}
where $C=2kLN_0\mu _{12}^{2}/{(\hbar\epsilon _0 {\gamma _2})}$ is a constant. Thus, the optimal value of $w^*$ is the maximum point of function $g(x,y,z,w)$ as Eq. (\ref{eq.g_func}).
\begin{figure*}
	\begin{equation}\label{eq.g_func}
		g\left( x,y,z,w \right) =\exp[-\frac{C z^2}{z^2+4w\left( x+y \right) ^2+2xz\left( x+y+z \right)}]\cdot \frac{z^{3/2}\left( x+y \right) \left( xz+4w\left( x+y \right) \right)}{\left[ z^2+4w\left( x+y \right) ^2+2xz\left( x+y+z \right) \right] ^2}.
	\end{equation}
\end{figure*}

Letting $u=x(x+y)$, and $v=z(2x+1)$, the optimal value for $w$ is
\begin{equation}
	w^*=\frac{x^2z\left( Cz-2u+\sqrt{4\left( u+v \right) ^2+\left( Cz \right) ^2} \right)}{8u^2}.
\end{equation}
The corresponding microwave field detuning is obtained as
\begin{equation}\label{eq.deltaL1}
	\Delta _{L}^{*}=\frac{\Omega _{p}^{2}\Omega _L}{2U}\sqrt{\frac{C\gamma _{2}^{2}\Omega _{L}^{2}-2U+\sqrt{4\left( U+V \right) ^2+\left( C\gamma _{2}^{2}\Omega _{L}^{2} \right) ^2}}{2}},
\end{equation}
where $U=\Omega _{p}^{2}\left( \Omega _{p}^{2}+\Omega _{c}^{2} \right)$, and $V=\Omega _{L}^{2}\left( 2\Omega _{p}^{2}+\gamma _{2}^{2} \right) $.

In high transmittance case, the optimization problem in Eq. (\ref{eq.opti1}) is simplified as follows.

\textbf{(P2): Sensitivity Maximization in High Transmittance Case via Local Microwave Detuning}
\begin{equation}\label{eq.opti2}
	\Delta _{L}^{**}=\underset{{\Delta _L}}{\text{argmax }} \chi _1\left( \Delta _L \right).
\end{equation}

According to Eq. (\ref{eq.chi1}) and Eq. (\ref{eq.sub}), we can get the following equivalent optimization problem,
\begin{equation}
	w^{**}=\underset{w}{\text{argmax }}\left\{ \begin{array}{c}
		\frac{z^{3/2}\left( x+y \right) \left( xz+4w\left( x+y \right) \right)}{\left[ z^2+4w\left( x+y \right) ^2+2xz\left( x+y+z \right) \right] ^2}\\
	\end{array} \right\}.
\end{equation}
Thus, the optimal value is the maximum point of function with respect to $w$,
\begin{equation}
	h \left( x,y,z,w \right) =\frac{z^{3/2}\left( x+y \right) \left( xz+4w\left( x+y \right) \right)}{\left[ z^2+4w\left( x+y \right) ^2+2xz\left( x+y+z \right) \right] ^2}.
\end{equation}
Deriving $h(x,y,z,w)$ with respect to $w$, we have
\begin{equation}
	\frac{\partial h}{\partial w}=\frac{4z^{3/2}\left( x+y \right) ^2\left( \left( 2x+1 \right) z^2-4w\left( x+y \right) ^2 \right)}{\left( 4w\left( x+y \right) ^2+z\left( 2x\left( x+y+z \right) +z \right) \right) ^3}.
\end{equation}
Since $x, y, z>0$, the optimal value for $w$ is
\begin{equation}
	w^{**}=\frac{\left( 2x+1 \right) z^2}{4\left( x+y \right) ^2},
\end{equation}
and the corresponding local microwave frequency detuning is
\begin{equation}\label{eq.deltaL2}
	\Delta _L^{**}=\frac{\Omega _{L}^{2}}{2\left( \Omega _{p}^{2}+\Omega _{c}^{2} \right)}\sqrt{2\Omega _{p}^{2}+\gamma _{2}^{2}}.
\end{equation}

\subsection{Probing Laser Detuning}
Assuming that $\Delta _c=\Delta _L=\gamma _3=\gamma _4=\gamma_c=\gamma=0$, the steady-state solution $\rho_{21}(\Delta_{p})$ can be obtained according to Eq. (\ref{eq.basic}) as Eq. (\ref{eq.rho21_p}).
\begin{figure*}
\begin{equation}\label{eq.rho21_p}
	\begin{aligned}
	\rho _{21}(\Delta_p) &=\frac{\Omega _p\left( 4\Delta _{p}^{2}-\Omega ^2 \right) \left( i\gamma _2\left( 4\Delta _{p}^{2}-\Omega ^2 \right) +2\Delta _p\left( 4\Delta _{p}^{2}-\Omega _{c}^{2}-\Omega ^2 \right) \right)}{D_p(\Omega)}, \\
	D_p(\Omega) &=64\Delta _{p}^{6}+\gamma _{2}^{2}\left( \Omega ^2-4\Delta _{p}^{2} \right) ^2 +4\Delta_{p}^{2}\left( \left( \Omega ^2+\Omega _{c}^{2} \right) ^2+2\Omega _{p}^{2}\left( \Omega _{p}^{2}+\Omega _{c}^{2}-2\Omega ^2 \right) \right) \\
	&-32\Delta _{p}^{4}\left( \Omega ^2+\Omega _{c}^{2}-\Omega _{p}^{2} \right) +2\Omega _{p}^{2}\Omega ^2\left( \Omega _{p}^{2}+\Omega _{c}^{2}+\Omega ^2 \right).
\end{aligned}
\end{equation}
\end{figure*}
In weak signal regime, $\Omega_s \ll \Omega_L$, the first order approximation of $\Omega_s / \Omega_L$ is given by Eq. (\ref{eq.rho21_p_Im}).
\begin{figure*}
\begin{equation}\label{eq.rho21_p_Im}
	\begin{aligned}
	\Im(\rho _{21}(\Delta_p))&\approx \frac{\gamma _2\Omega _p\left( 4\Delta _{p}^{2}-\Omega _{L}^{2} \right) ^2}{D_p(\Omega_{L})}-\frac{4\gamma _2\Omega _p\Omega _L A_p \Omega _s\cos  S\left( t \right)}{D_p^2(\Omega_{L})},\\
	A_p&=\left( 4\Delta _{p}^{2}-\Omega _{L}^{2} \right) \left( \Omega _{L}^{2}\Omega _{p}^{2}\left( \Omega _{c}^{2}+\Omega _{p}^{2} \right) -16\Delta _{p}^{4}\Omega _{c}^{2}+4\Delta _{p}^{2}\left( \Omega _{c}^{2}\left( \Omega _{c}^{2}+\Omega _{L}^{2} \right) \right) \right)\\
	& +3\Omega _{p}^{2}\left( 4\Delta _{p}^{2}-\Omega _{L}^{2} \right)\left( \Omega _{p}^{2}+\Omega _{c}^{2} \right).\\
	\end{aligned}
\end{equation}
\end{figure*}
The corresponding imaginary component of linear susceptibility $\chi \left( t, \Delta_p \right)$ of the atomic system can be written as
\begin{equation}
	\Im\left( \chi \left( t, \Delta_p \right) \right) =\chi _{0}(\Delta_p)+\chi _{1}(\Delta_p)\Omega _s\cos S(t),
\end{equation}
which has a time-invariant part 
\begin{equation}
		\chi _{0}(\Delta_p)=-\frac{2N_0\mu _{12}^{2}}{\epsilon _0{\hbar }\Omega _p} \frac{\gamma _2\Omega _p\left( 4\Delta _{p}^{2}-\Omega _{L}^{2} \right) ^2}{D_p(\Omega_{L})},\\
\end{equation}
and a time variant part
\begin{equation}
	\chi _{1}(\Delta_p)=\frac{2N_0\mu _{12}^{2}}{\epsilon _0{\hbar }\Omega _p}\frac{4\gamma _2\Omega _p\Omega _LA_p}{D_p^2(\Omega_{L})}.
\end{equation}

The sensitivity optimization problem in the general case is formulated as follows.

\textbf{(P3): Sensitivity Maximization in General Case via Probing Laser Detuning}
\begin{equation}
	\Delta _{p}^{*}=\underset{{\Delta _p}}{\text{argmax }}\kappa(\Delta_p) =\underset{{\Delta _p}}{\text{argmax }} e^{-kL\chi _0\left( \Delta _p \right)}\chi _1\left( \Delta _p \right).
\end{equation}

We numerically characterize the probing laser detuning effect on the weak signal response in Section IV. 

\subsection{Coupling Laser Detuning}

Assuming that $\Delta _p=\Delta _L=\gamma _3=\gamma _4=\gamma_c=\gamma=0$, the steady-state solution $\rho_{21}(\Delta_{c})$ can be obtained according to Eq. (\ref{eq.basic}) as Eq. (\ref{eq.rho21_c}).
\begin{figure*}
	\begin{equation}\label{eq.rho21_c}
	\begin{aligned}
		\rho _{21}(\Delta_c) &=\frac{i\gamma _2\Omega _p\left( 4\Delta _{c}^{2}-\Omega ^2 \right) ^2-2\Omega _p\Omega _{c}^{2}\Delta _c\left( 4\Delta _{c}^{2}-\Omega ^2 \right)}{D_c(\Omega)}, \\
		D_c(\Omega) &=32\Delta _{c}^{4}\Omega _{p}^{2}+\gamma _{2}^{2}\left( \Omega ^2-4\Delta _{c}^{2} \right) ^2+2\Omega _{p}^{2}\Omega ^2\left( \Omega _{p}^{2}+\Omega _{c}^{2}+\Omega ^2 \right) \\
		& +4\Delta _{c}^{2}\left( \left( \Omega _{p}^{2}+\Omega _{c}^{2} \right) ^2+\Omega _{p}^{2}\left( \Omega _{p}^{2}-4\Omega ^2 \right) \right). 
	\end{aligned}
\end{equation}
\end{figure*}
In weak signal regime, $\Omega_s \ll \Omega_L$, the first order approximation of $\Omega_s / \Omega_L$ is given by Eq. (\ref{eq.rho21_c_Im}).
\begin{figure*}
	\begin{equation}\label{eq.rho21_c_Im}
	\begin{aligned}
		\Im(\rho _{21}(\Delta_c))&\approx \frac{\gamma _2\Omega _p\left( 4\Delta _{c}^{2}-\Omega _{L}^{2} \right) ^2}{D_c(\Omega_{L})}+\frac{4\gamma _2\Omega _p\Omega _LA_c\Omega _s\cos \left( S\left( t \right) \right)}{D_c^2(\Omega_{L})},\\
		A_c&=\left( 4\Delta _{c}^{2}-\Omega _{L}^{2} \right) \left( \Omega _{L}^{2}\Omega _{p}^{2}\left( \Omega _{p}^{2}+\Omega _{c}^{2} \right) +4\Delta _{c}^{2}\left( \Omega _{c}^{4}+3\Omega _{c}^{2}\Omega _{p}^{2}+3\Omega _{p}^{4} \right) \right).
	\end{aligned}
\end{equation}
\end{figure*}
The corresponding imaginary component of linear susceptibility $\chi \left( t, \Delta_c \right)$ of the atomic system can be written as
\begin{equation}
	\Im\left( \chi \left( t, \Delta_c \right) \right) =\chi _{0}(\Delta_c)+\chi _{1}(\Delta_c)\Omega _s\cos S(t),
\end{equation}
which has a time-invariant part 
\begin{equation}
	\chi _{0}(\Delta_c)=-\frac{2N_0\mu _{12}^{2}}{\epsilon _0{\hbar }\Omega _p} \frac{\gamma _2\Omega _p\left( 4\Delta _{c}^{2}-\Omega _{L}^{2} \right) ^2}{D_c(\Omega_{L})},\\
\end{equation}
and a time variant part
\begin{equation}
	\chi _{1}(\Delta_c)=-\frac{2N_0\mu _{12}^{2}}{\epsilon _0{\hbar }\Omega _p}\frac{4\gamma _2\Omega _p\Omega _LA_c}{D_c^2(\Omega_{L})}.
\end{equation}

The sensitivity optimization problem in the general case is formulated as follows.

\textbf{(P4): Sensitivity Maximization in General Case via Coupling Laser Detuning}
\begin{equation}
	\Delta _{c}^{*}=\underset{{\Delta _c}}{\text{argmax }}\kappa(\Delta_c) = \underset{{\Delta _c}}{\text{argmax }} e^{-kL\chi _0\left( \Delta _c \right)}\chi _1\left( \Delta _c \right) .
\end{equation}

We also numerically characterize the coupling laser detuning effect on the weak signal response in Section IV. 
 
\subsection{Transit Relaxation Effect}
\begin{figure}[htbp]
	\centering
	\includegraphics[width = 3.5in]{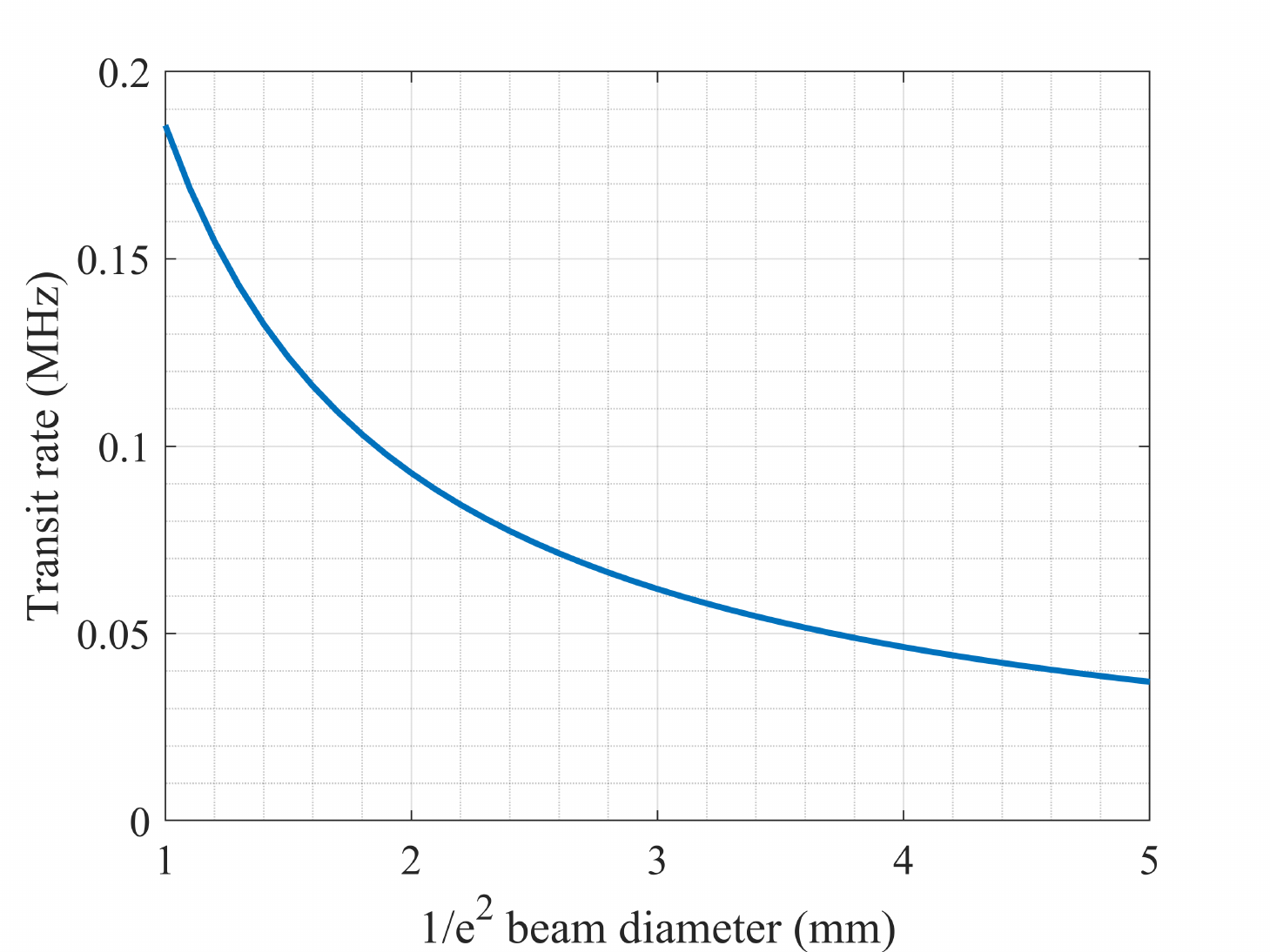}
	\caption{Transit relaxation rate to the $1/e^2$ beam waist at $T=300$ K.}
	\label{fig.gamma}
\end{figure}

Our approach adpots each non-interacting and independent atom that participates as an identical and stable atomic sensor of the microwave electric field. Except for spontaneous radiation and atoms collision, transit relaxation is an important type of relaxation to thermal vapor Rydberg atomic sensors. Due to random motion of atoms, atoms continuously enter or leave the area covered by the laser beam. We assume that atoms in the ground state enter the interaction region from outside the beam at a rate $\gamma$, while the atoms in different states in the interaction region leave at the same rate. When the atoms leave, the number of excited atoms and the coherence between the different states are destroyed. The characteristic time for this process to occur is determined by the average time of flight through the cross section of the laser beams, which defines the effective relaxation rate [16]
\begin{equation}
	\gamma =\sqrt{\frac{8k_BT}{\pi m}}\frac{1}{w\sqrt{2\ln 2}},
\end{equation}
where $w$ is the $1/e^2$ beam waist, $m$ is the atom mass, $T$ the ensemble temperature, and $k_B$ is Boltzmann’s constant. Fig. \ref{fig.gamma} shows the relationship between the $1/e^2$ beam waist and transit relaxation rate for Cs atoms at $T=300$ K.

For non-zero transit relaxation rate $\gamma$, we assume that $\Delta _p=\Delta _L=\Delta_p=\gamma _3=\gamma_4=\gamma_c=0$. We also expect that the corresponding imaginary component of linear susceptibility has a time-invariant part and a time variant part,
\begin{equation}
	\Im\left( \chi \left( t, \gamma \right) \right) =\chi _{0}(\gamma)+\chi _{1}(\gamma)\Omega _s\cos \left(S(t) \right).
\end{equation}

The sensitivity optimization problem is formulated as follows.

\textbf{(P5): Sensitivity Maximization in General Case via Transit Relaxation Rate}
\begin{equation}\label{eq.optig}
	\gamma^{*}=\underset{{\gamma}}{\text{argmax }}\kappa(\gamma) = e^{-kL\chi _0\left( \gamma \right)}\chi _1\left( \gamma \right).
\end{equation}

We numerically characterize the transit relaxation effect on the weak signal response in Section IV 

\section{Numerical Results}\label{sec.numerical}
We adopt some parameters from the experimental setups in \cite{jing2020atomic}. Cs atoms were filled in a vapour cell at room temperature. The cell contained ground-state atoms at a total density of $N_0=4.89 \times 10^{10}$ cm$^{-3}$. For Rydberg atom EIT, the residence time in the Rydberg state is small. Under typical conditions, the population in Rydberg states is $~0.0001$. Only a small distribution of atomic velocity classes in the vapor cell are selected by the EIT lasers, roughly $~1/400$ for the scheme shown in Fig. \ref{fig.level} \cite{fan2015atom}. The effective atoms density for atomic vapor is approximated as $N_{0,\text{eff}}\approx 10^{8}$ cm$^{-3}$. The four-level configuration in Fig. \ref{fig.level} using four states in a Cs atom: $|1\rangle \rightarrow 6S_{1/2}, F=4$; $|2\rangle \rightarrow 6P_{3/2}, F=5$; $|3\rangle \rightarrow 47D_{5/2}$; and $|4\rangle \rightarrow 48P_{3/2}$. States $|2\rangle$, $|3\rangle$, and $|4\rangle$ have the inverse lifetimes $\gamma_2=2\pi\times 5.2$ MHz, $\gamma_3=2\pi \times 3.9$ kHz, and $\gamma_{4}=2\pi\times 1.7$ kHz. In this section, we ignore the atoms collision effect, $\gamma_c=0$, the effective Rabi frequencies for the transitions $|1\rangle \rightarrow |2\rangle$ and $|2\rangle \rightarrow |3\rangle$ are $\Omega_p=2\pi\times 5.7$ MHz, $\Omega_c=2\pi\times 0.97$ MHz, respectively. The frequency intervals between the hyperfine states $6P_{3/2}$ are $151.2$ MHz ($F=2\rightarrow F=3$), $201.2$ MHz ($F=3\rightarrow F=4$), and $251.1$ MHz ($F=4\rightarrow F=5$) \cite{steck2003cesium}. In the case of resonance, the system is locked to a specific hyperfine state. We consider a detuning range from $-50$ MHz to $50$ MHz in numerical results. 

\begin{figure}[htbp]
	\centering
	\includegraphics[width = 3.5 in]{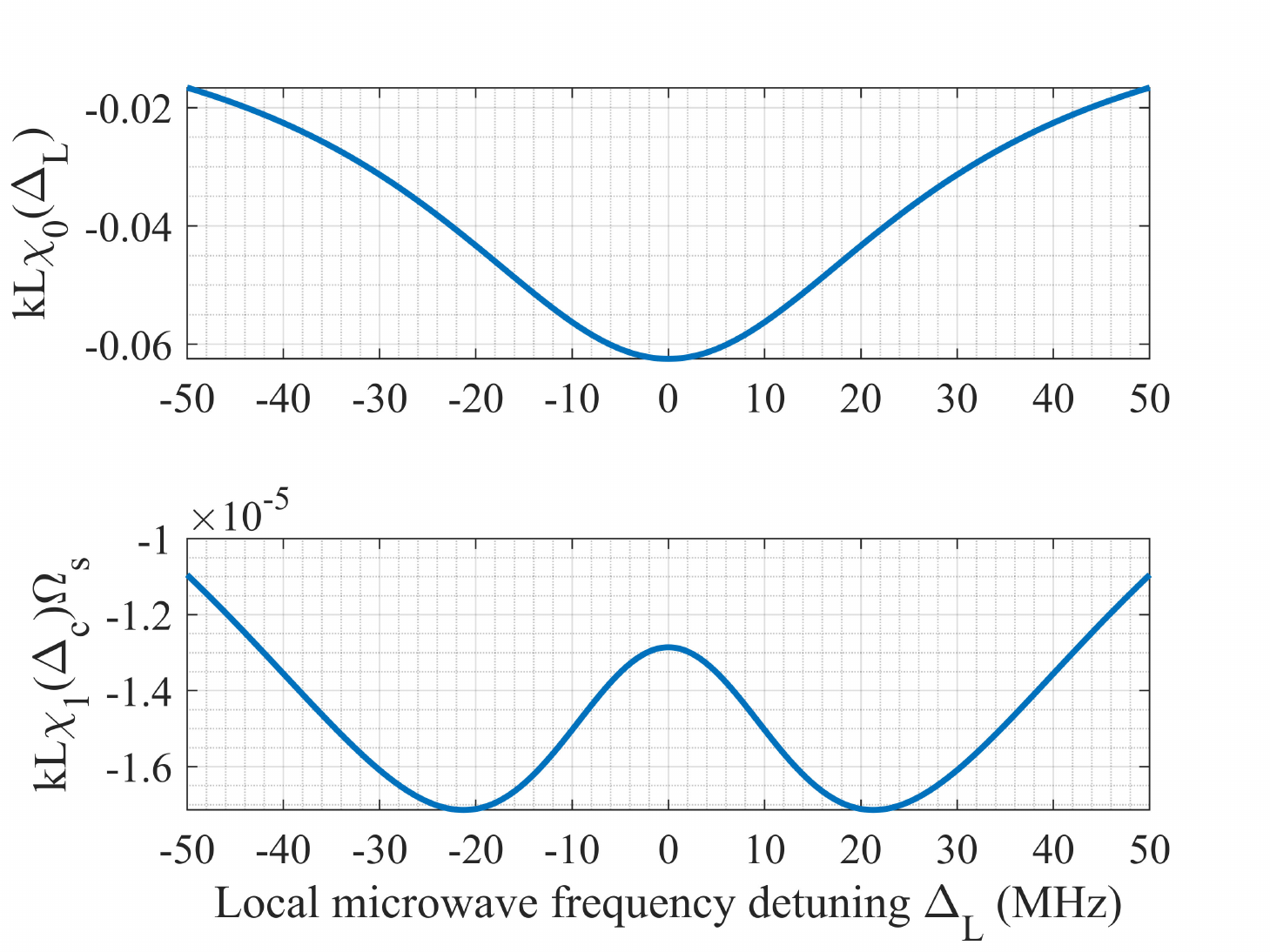}
	\caption{The conversion coefficient $kL\chi_{0}(\Delta_{L})$ and $kL\chi_{1}(\Delta_{L})\Omega_s$ with respect to local microwave detuning $\Delta_{L}$ ($\Omega_s=2\pi \times 1$ kHz, $L=1$ cm).}
	\label{fig.l condition}
\end{figure}

In Section III, we assume that $|kL\chi_1(\Delta)\Omega_s| \ll 1$ and solve the optimization problem \textbf{(P1)} to maximize the conversion coefficient $|\kappa(\Delta)|$. In high transmittance case, the optimization problem is simplified as \textbf{(P2)} to maximize $|\kappa '(\Delta)|$. According to Eq. (\ref{eq.steady}), the valid range for the general case is estimated as $\Omega_s < 10$ kHz. The high transmittance case corresponds to thin medium with $L \le 1$ cm. For $L=1$ cm and $\Omega_{s}=2\pi\times 1$ kHz, the numerical values of $kL\chi_{0}(\Delta_{L})$ and $kL\chi_{1}(\Delta_{L})\Omega_s$ are shown in Fig. \ref{fig.l condition}, which means that the assumptions in general case and high transmittance case can be satisfied.

\begin{figure}[htbp]
	\centering
	\includegraphics[width = 3.5 in]{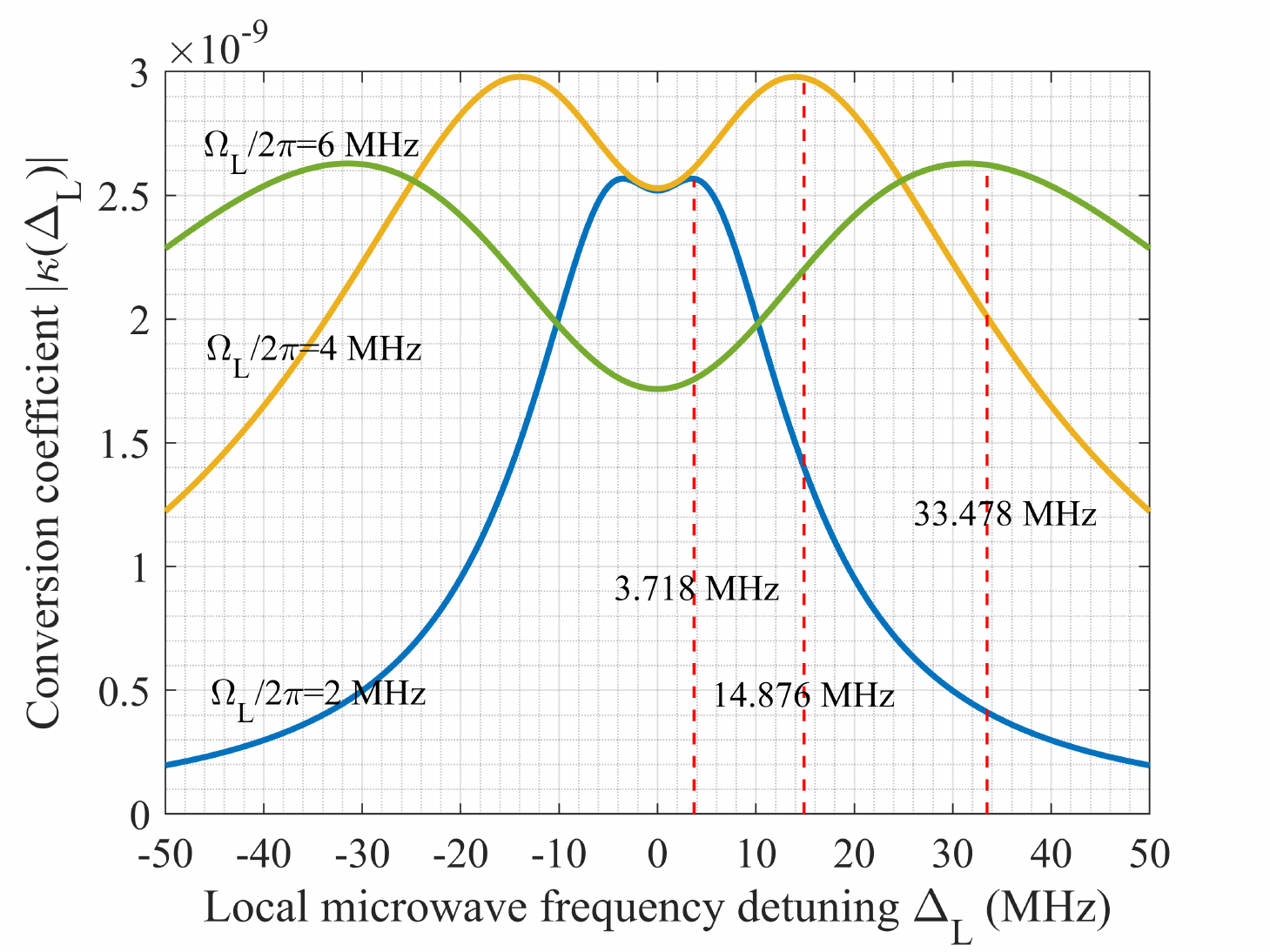}
	\caption{Conversion coefficient $|\kappa(\Delta_L)|$ with respect to local microwave detuning $\Delta_{L}$ in the general case \textbf{(P1)}.}
	\label{fig.optimal l1}
\end{figure}

\begin{figure}[htbp]
	\centering
	\includegraphics[width = 3.5 in]{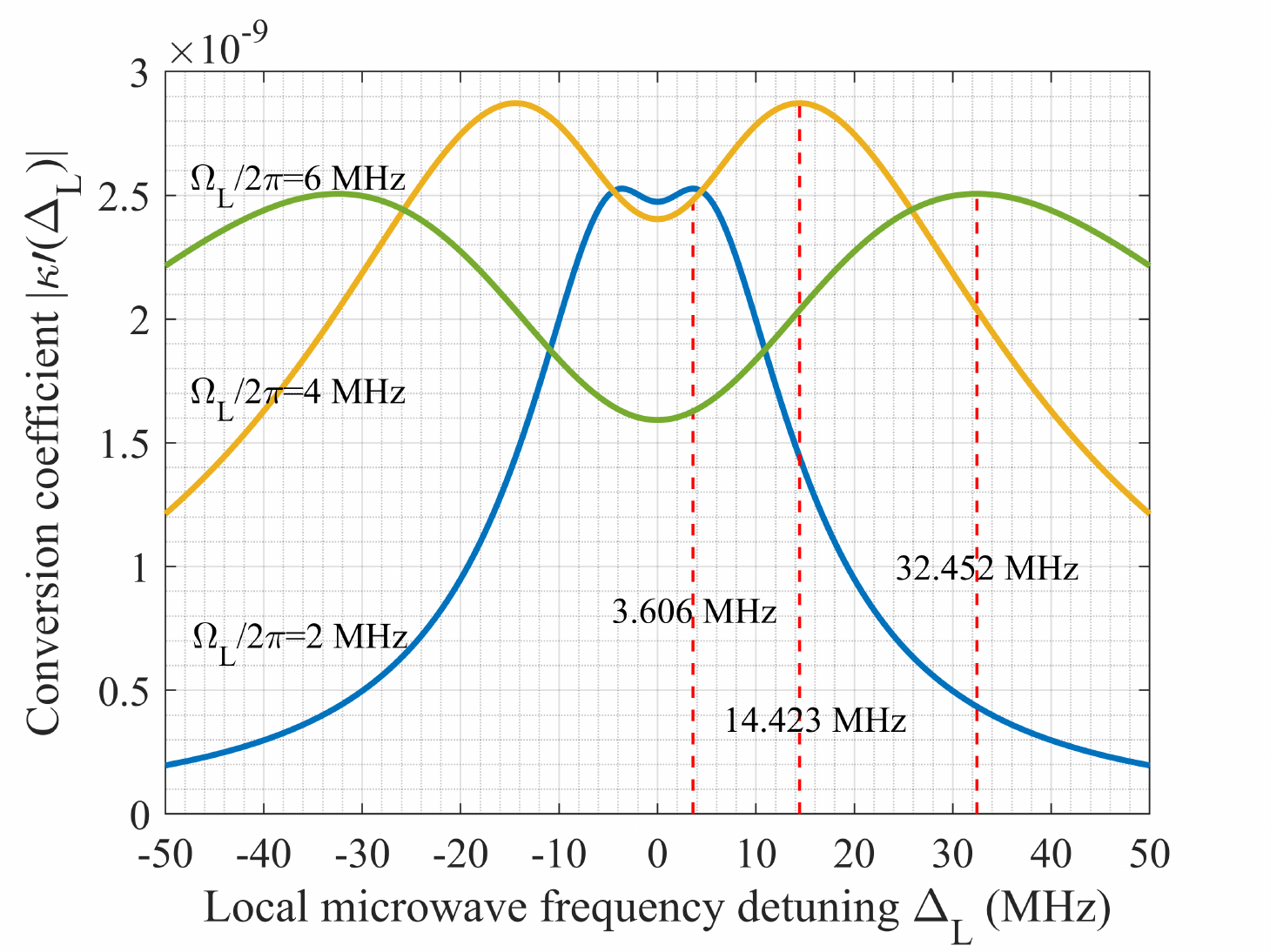}
	\caption{Conversion coefficient $|\kappa '(\Delta_L)|$ with respect to local microwave detuning $\Delta_{L}$ in high transmittance case \textbf{(P2)}.}
	\label{fig.optimal l2}
\end{figure}

For local microwave frequency detuning, the conversion coefficient $|\kappa(\Delta_L)|$ in the general case is shown in Fig. \ref{fig.optimal l1}, and conversion coefficient $|\kappa '(\Delta_L)|$ in high transmittance case is shown in Fig. \ref{fig.optimal l2}. According to Eq. (\ref{eq.deltaL1}) and Eq. (\ref{eq.deltaL2}), the theoretical optimal values are marked as dashed red lines. It is seen that the solution in high transmittance case is numerically close to the original one in the general case. In the general case, the sensitivity gains at $\Omega_{L}/2\pi=2$ MHz, 4 MHz, 6 MHz by local microwave frequency detuning are 0.06 dB, 0.71 dB, and 1.85 dB, respectively. In high transmittance case, the sensitivity gains at $\Omega_{L}/2\pi=2$ MHz, 4 MHz, 6 MHz by local microwave frequency detuning are 0.09 dB, 0.77 dB, and 1.96 dB, respectively. It is predictable that stronger local microwave power can result in greater sensitivity gains.

\begin{figure}[htbp]
	\centering
	\includegraphics[width = 3.5 in]{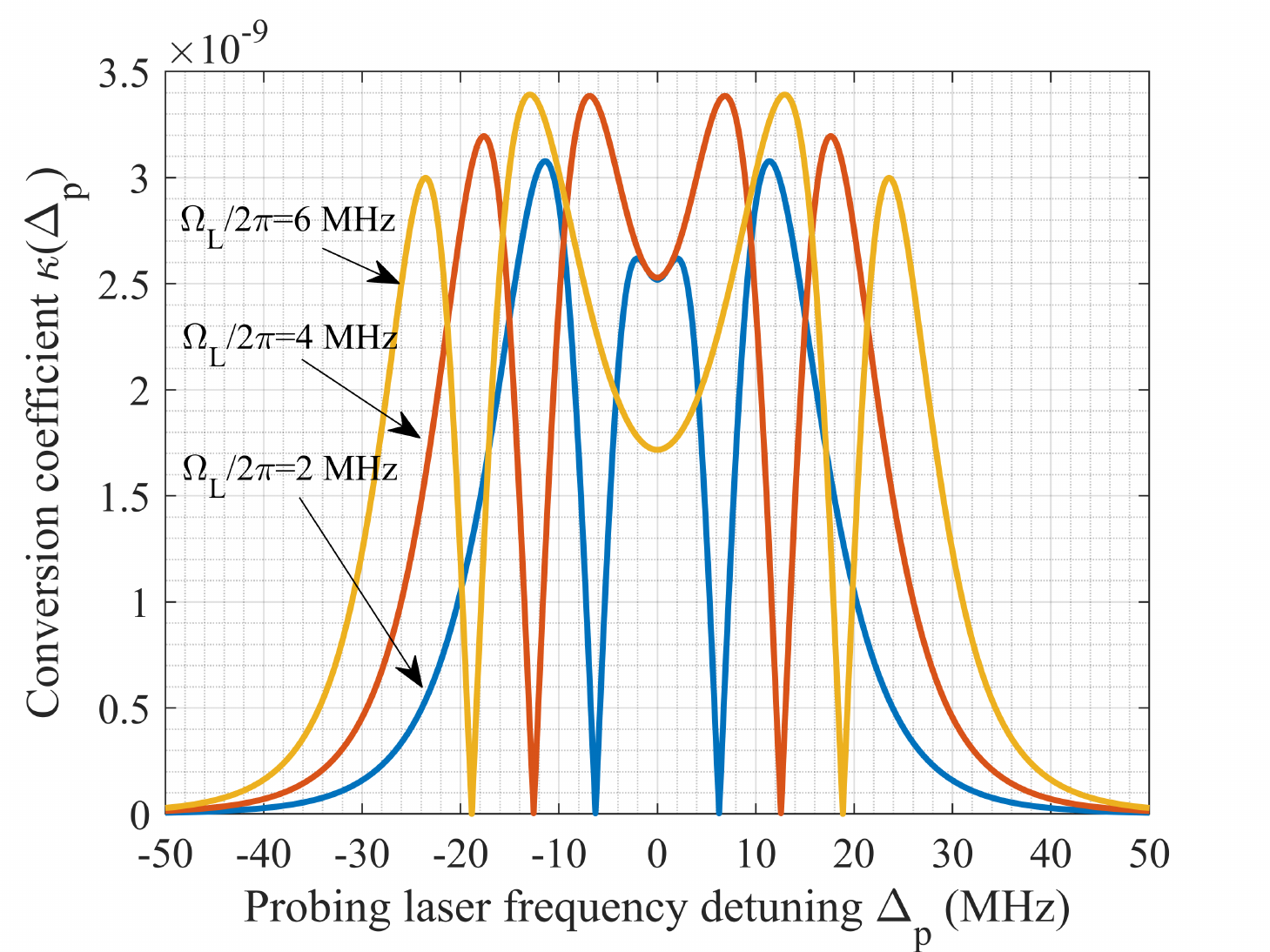}
	\caption{Conversion coefficient $|\kappa(\Delta_p)|$ with respect to probing laser detuning $\Delta_{p}$ in the general case \textbf{(P3)}.}
	\label{fig.optimal p}
\end{figure}

\begin{figure}[htbp]
	\centering
	\includegraphics[width = 3.5 in]{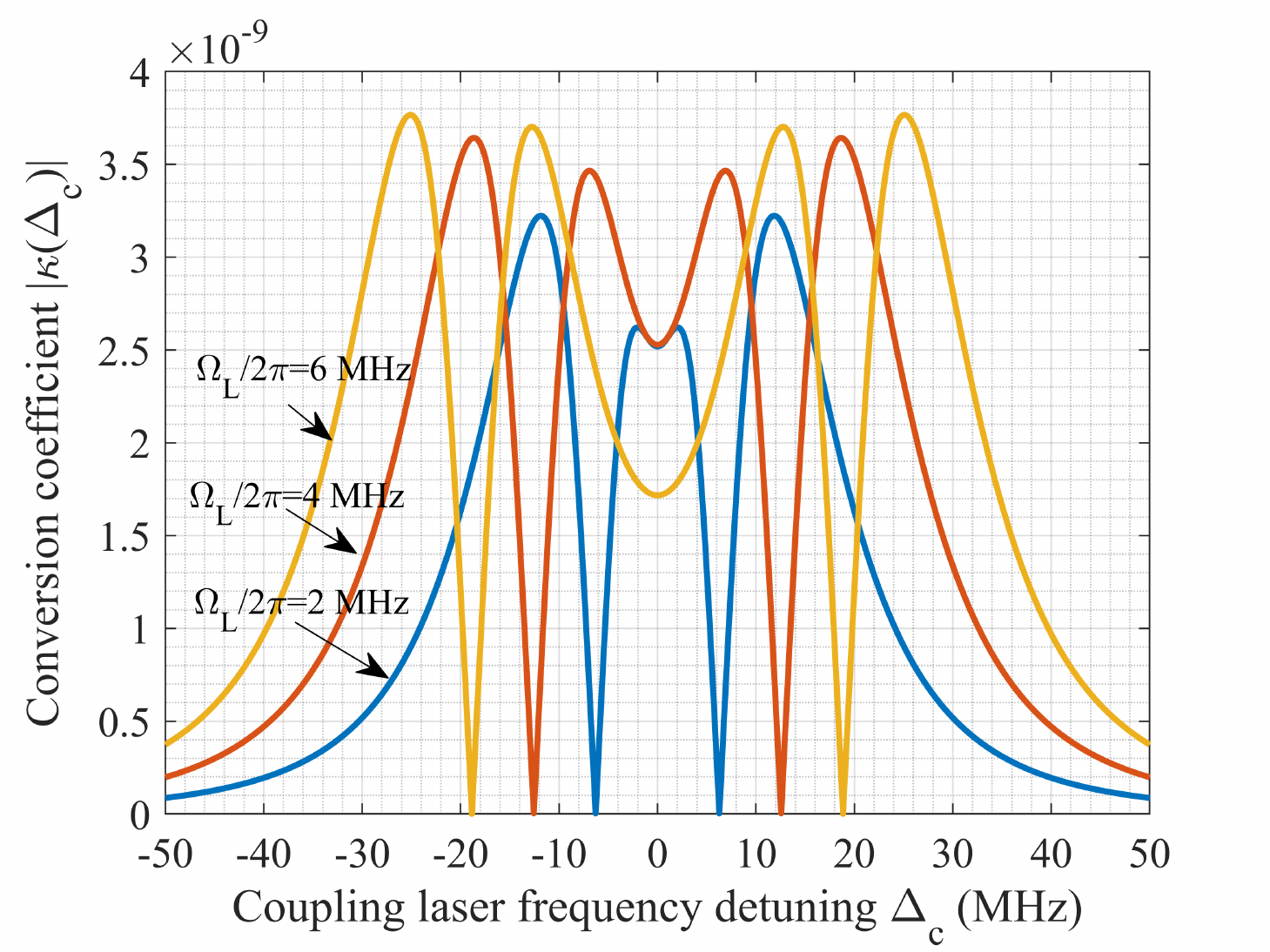}
	\caption{Conversion coefficient $|\kappa(\Delta_c)|$ with respect to coupling laser detuning $\Delta_{c}$ in the general case \textbf{(P4)}.}
	\label{fig.optimal c}
\end{figure}

For probing and coupling laser frequency detunings, conversion coefficients $|\kappa(\Delta_p)|$ and $|\kappa(\Delta_c)|$ in the general case are shown in Fig. \ref{fig.optimal p} and Fig. \ref{fig.optimal c}, respectively. The ideal optimization problem is to maximize the absolute value of $\kappa(\Delta_p)$ and $\kappa (\Delta_c)$, but the analytical optimal values are obtained by the normal form. Thus, we can find another local maximal value, corresponding to the minimal value of $\kappa(\Delta_p)$ and $\kappa (\Delta_c)$. In the general case, the sensitivity gains at $\Omega_{L}/2\pi=2$ MHz, 4 MHz, 6 MHz by probing laser frequency detuning are 0.85 dB, 1.27 dB, and 2.96 dB, respectively. And the sensitivity gains at $\Omega_{L}/2\pi=2$ MHz, 4 MHz, 6 MHz by coupling laser frequency detuning are 1.05 dB, 1.59 dB, and 3.41 dB, respectively.

\begin{figure}[htbp]
	\centering
	\includegraphics[width = 3.5 in]{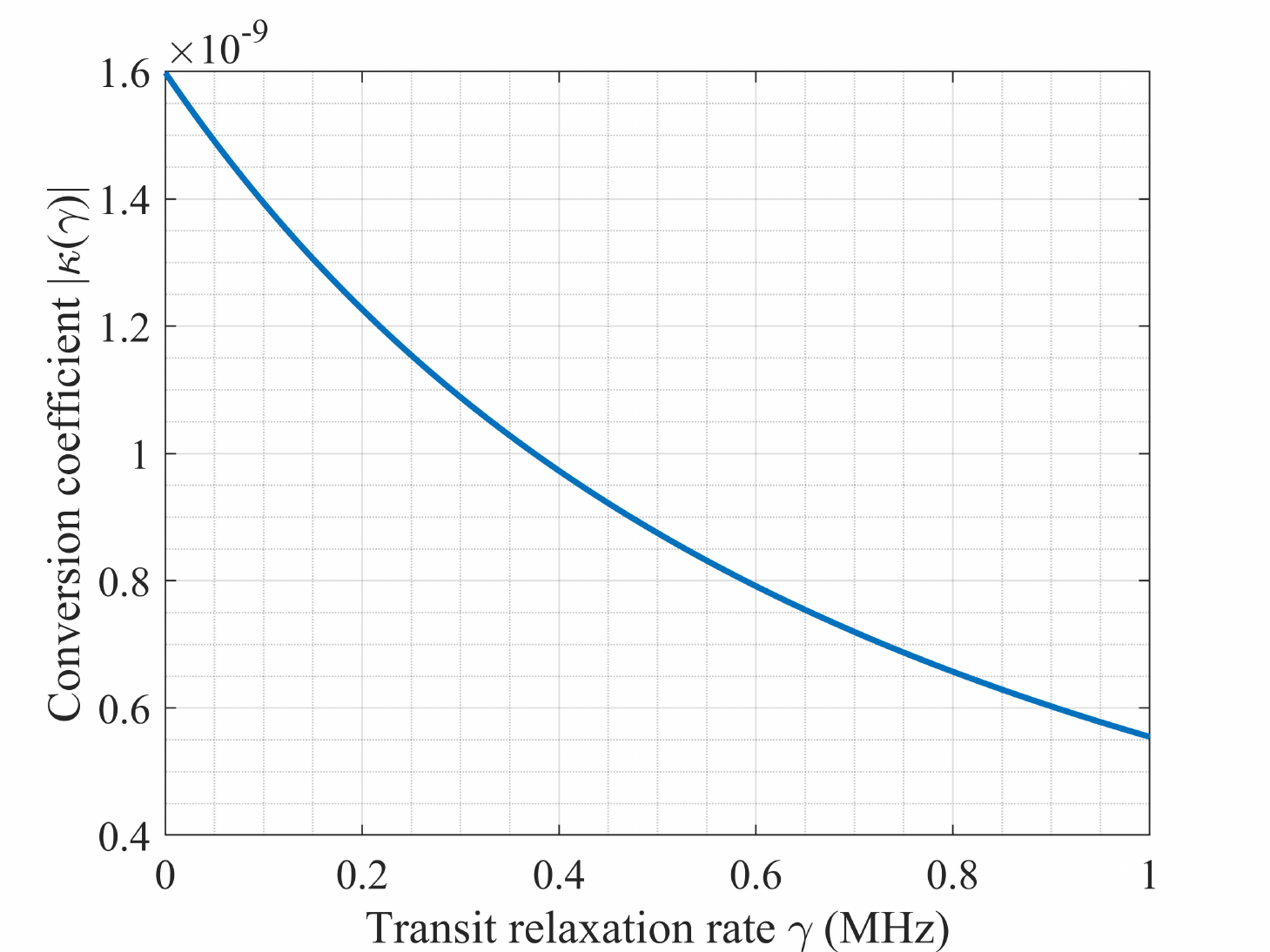}
	\caption{Conversion coefficient $|\kappa(\gamma)|$ with respect to transit relaxation rate $\gamma$ in the general case \textbf{(P5)}.}
	\label{fig.g}
\end{figure}

As the transit relaxation rate increases, the conversion coefficient decreases, as shown in Fig. \ref{fig.g}. We can reduce the influence of the transit relaxation effect by reducing the size of the atomic cavity and the atomic number density. Correspondingly, the number of effective atoms involved in the interaction decreases, and the detection sensitivity reduces.

\section{Conclusion}
Based on the atomic four-level model, we have derived the steady-state solutions of the density matrix of the Rydberg atomic system operating with frequency detuning. Given the input microwave signal, we have established an optimization problem of the linear susceptibility of atomic medium to maximize the amplitude of the output signal. We have theoretically obtained the optimal value of local microwave electric field frequency detuning and numerically analyzed the effects of probing laser and coupling laser frequency detuning on the optimization target. In addition, considering the existence of the atomic transit relaxation effect, we have modified the system Liouville equation, and derived and shown the influence of the transit relaxation rate on the system detection sensitivity and the time to reach steady state. Such results show that setting the appropriate frequency detuning can improve the system detection sensitivity, resulting in extra several dB sensitivity gain through this method when the device performance cannot be further improved. Considering the relationship between the local microwave electric field and the microwave signal in the heterodyne scheme, such results also mean that the heterodyne Rydberg atomic receiver can provide the same or better sensitivity as the resonance case in continuous weak microwave signal detection over a large frequency range \cite{hu2022continuous}. The relaxation effect reduces the sensitivity of the system, which should be eliminated or reduced in experiments.

\bibliographystyle{./IEEEtran}
\bibliography{./mybib}

\end{document}